\documentclass[conference,a4paper]{IEEEtran}

\usepackage{amsmath, amsthm, amssymb}
\usepackage{amsfonts}
\usepackage{amsthm}
\usepackage{tikz}
\usetikzlibrary{shapes,arrows}
\usepackage{verbatim}
\usepackage[]{subfigure}
\theoremstyle{remark}

\usepackage{graphicx,relsize,verbatim,enumerate,xcolor}
\usepackage{pgfplots}
\usetikzlibrary{arrows,positioning,fit,backgrounds,shapes}
\usetikzlibrary{calc}

\newtheorem{theorem}{Theorem}
\newtheorem{definition}{Definition}

\tikzstyle{arw}=[->,>=latex]
\tikzstyle{node}=[rectangle,draw,outer sep=0pt,minimum width=1.7cm, minimum height=8mm]

\newcommand{\E}{{\mathbb{E}}}
\newcommand{\br}[1]{{\left(#1\right)}}
\newcommand{\setbr}[1]{{\left\{#1\right\}}}
\newcommand{\abs}[1]{{\left|#1\right|}}

\newcommand{\sq}[1]{{\left[#1\right]}}
\newcommand{\R}{{\mathbb{R}}}

\newcommand{\defeq}{{~\triangleq~}}

\newcommand{\x}{{\hat{x}}}

\newcommand{\inv}{{^{-1}}}

\newcommand{\D}{{\Delta}}

\newcommand{\Rcal}{\mathcal{R}}

\newcommand{\Xcal}{\mathcal{X}}
\newcommand{\Xh}{\hat{\mathcal{X}}}
\newcommand{\Ycal}{\mathcal{Y}}
\newcommand{\Zcal}{\mathcal{Z}}
\newcommand{\X}{{\hat{X}}}

\newcommand{\bp}{{\bar{P}}}
\newcommand{\bpa}{{\bar{P}^{(1)}}}

\newcommand{\As}{\textsf{A}~}
\newcommand{\Bs}{\textsf{B}~}
\newcommand{\Hs}{\textsf{H}~}
\newcommand{\Es}{\textsf{E}~}

\newcommand{\A}{\textsf{A}}
\newcommand{\B}{\textsf{B}}
\newcommand{\He}{\textsf{H}}
\newcommand{\Ee}{\textsf{E}}

\newcommand{\s}{{\mathcal{S}}}

\newcommand{\es}{{\emptyset}}

\newcommand{\setb}[1]{{\left\{#1\right\}}}

\newcommand{\normtv}[1]{{\left\|#1\right\|_{\mbox{\tiny{$TV$}}}}}

\begin{document}

\sloppy

%% Paper Title
%% You can use linebreaks \\ within to get better formatting as
%% desired. 
\title{Secure Coordination with a Two-Sided Helper} 

\author{
  \IEEEauthorblockN{Sanket Satpathy and Paul Cuff}
  \IEEEauthorblockA{Dept. of Electrical Engineering\\
    Princeton University\\
    Princeton, USA\\
    Email: \{satpathy,cuff\}@princeton.edu}
}

%% Create the title:
\maketitle

%% Abstract: 
%% For the final version of the accepted paper, please make sure you
%% remove the comment "THIS PAPER IS ELIGIBLE FOR THE STUDENT PAPER
%% AWARD."
%%
\begin{abstract}
  We investigate the problem of secure source coding with a two-sided helper in a game-theoretic framework. Alice (\textsf{A}) and Helen (\textsf{H}) view iid correlated information sequences $X^n$ and $Y^n$ respectively. Alice communicates to Bob (\textsf{B}) at rate $R$, while \Hs broadcasts a message to both \As and \Bs at rate $R_H$. Additionally, \As and \Bs share secret key $K$ at rate $R_0$ that is independent of $(X^n,Y^n)$. An active adversary, Eve (\textsf{E}) sees all communication links while having access to a (possibly degraded) version of the past information. 
  We characterize the rate-payoff region for this problem. We also solve the problem when the link from \As to \Bs is private. Our work recovers previous results of Schieler-Cuff and Kittichokechai et al.
%  Switching the roles of \As and \Hs yields a result on triangular source coding.
%  We also resolve a result claimed by Kaspi-Berger and refuted by Permuter et al.
\end{abstract}

%\begin{abstract}
%  [To be considered for an IEEE Jack Keil Wolf ISIT Student Paper Award.] We investigate the problem of secure source coding with a two-sided helper in a game-theoretic framework. Players \textsf{A} and \textsf{H} view iid correlated information sequences $X^n$ and $Y^n$ respectively. Player \As communicates to \textsf{B} at rate $R$, while \Hs broadcasts a message to both \As and \Bs at rate $R_H$. Additionally, \As and \Bs share secret key $K$ at rate $R_0$ that is independent of $(X^n,Y^n)$. An active adversary, \textsf{E} sees all communication links while having access to a (possibly degraded) version of the past information. 
%  We characterize the rate-payoff region for this problem. We also solve the problem when the link from \As to \Bs is private. Our work recovers previous results of Schieler-Cuff and Kittichokechai et al. Switching the roles of \As and \Hs yields a result on triangular source coding. We also resolve a result claimed by Kaspi-Berger and refuted by Permuter et al.
%\end{abstract}

\section{Introduction}

There has been significant recent interest in secure source coding \cite{vp,gund,tand,kc1,kc2,Cuff3,rdss}. Settings involving secret key and helpers have been studied. Most of these approaches to secrecy consider distortion at the legitimate receiver, and equivocation (equivalently, information leakage) at the eavesdropper. As such, they forsake an intrinsic allure of information theory results. Shannon's information measures are used in the problem formulation, rather than appearing as the answer to a purely operational question.

Of course, it would be wrong to say that an equivocation-based approach has no operational implication. As Wyner \cite{wire} notes, high equivocation would imply a high probability of error if the eavesdropper tried to reconstruct the entire message block. The extremes of equivocation correspond to perfect secrecy and error-free decoding. Both these cases can be defined by simple operational statements.

Recently, Cuff \cite{Cuff3,Cuff4,rdss} proposed a distortion-based approach to secrecy in which the past information is causally revealed to the eavesdropper. This formulation of partial secrecy is natural when understood in a game-theoretic context. A repeated zero-sum game is being played by the adversary versus the communication system. Distortion is now replaced by payoff, while the information sequences equate to actions of the players.  Settings of distributed control \cite{bork} can be viewed as a repeated zero-sum game.

Remarkably, when the payoff is chosen to be the log-loss function \cite{court}, the above framework recovers results for (normalized) equivocation-based secrecy \cite{eq}. Under this choice of payoff, the adversary expresses her belief about the distribution of the information sequence. Additionally, applications of log-loss to the study of information bottleneck \cite{bottle} and image processing \cite{gray} have been explored.

\tikzstyle{block} = [draw, fill=blue!20, rectangle, 
    minimum height=2em, minimum width=4em]
\tikzstyle{sum} = [draw, fill=blue!20, circle, node distance=1cm]
\tikzstyle{input} = [coordinate]
\tikzstyle{output} = [coordinate]
\tikzstyle{pinstyle} = [pin edge={to-,thin,black}]

\begin{figure}[h]
\begin{center}
 \resizebox{3.4in}{!}{\begin{tikzpicture}
 [node distance=1cm,minimum width=1cm,minimum height =.75 cm]
  \node[rectangle,minimum width=5mm] (source) {$X^n$};
  \node[node, fill=blue!20] (alice) [right =7mm of source] {\As};
  \node[node, fill=blue!20] (bob) [right =4cm of alice] {\Bs};
  \node[coordinate] (dummy) at ($(alice.east)!0.5!(bob.west)$) {};
  \node[rectangle,minimum width=5mm] (xhat) [right =7mm of bob] {$\hat{X}^n$};
  \node[rectangle,minimum width=7mm] (key) [above =7mm of dummy] {$\textcolor{blue}{M_H},K\in[2^{nR_0}]$};
  \node[node, fill=red!20] (eve) [below =5mm of bob] {\Es};
  \node[rectangle,minimum width=5mm] (zn) [right =7mm of eve] {$Z^n$};
  \node[rectangle] (side) [below=5mm of eve] {$(M,\textcolor{blue}{M_H},D^{i-1})$};
  
  \node[rectangle,minimum width=5mm] (xinput) at (source.south east |- side.west) {$Y^n$};
  \node[node, fill=blue!20] (chx) [right =8.5mm of xinput] {\Hs};
  \node[rectangle,minimum width=5mm] (xoutput) [right =5mm of chx] {$\textcolor{blue}{M_H}\in [2^{n R_H}]$};
  
%  \node[rectangle,minimum width=5mm] (yinput) [below =3mm of xinput] {$(X_i,Y_i)$};
%  \node[node] (chy) [right =5mm of yinput] {$P_{D|X,Y}$};
%  \node[rectangle,minimum width=5mm] (youtput) [right =5mm of chy] {$D_i$};

  \draw [arw] (source) to (alice);
  \draw [arw] (alice) to node[minimum height=6mm,inner sep=0pt,midway,above]{$M\in[2^{nR}]$} (bob);
  \draw [arw] (bob) to (xhat);
  \draw [arw] (key) to [out=180,in=90] (alice);
  \draw [arw] (key) to [out=0,in=90] (bob);
%  \draw [arw,rounded corners] (dummy) |- (eve);
  \draw [arw] (eve) to (zn);
  \draw [arw] (side) to (eve);
  \draw [arw] (xinput) to (chx);
  \draw [arw] (chx) to (xoutput);
%  \draw [arw] (yinput) to (chy);
%  \draw [arw] (chy) to (youtput);
  %\draw [draw,dashed,gray] (xyinput.south west) rectangle (output.north east);
 \end{tikzpicture}}
 \caption{\small Causal-disclosure secrecy with a two-sided helper.} %We allow disclosure $D=(X,D_{x})$ with $P_{D_{x}|X}$ arbitrary.}
 \label{pstsh}
 \end{center}
 \end{figure}
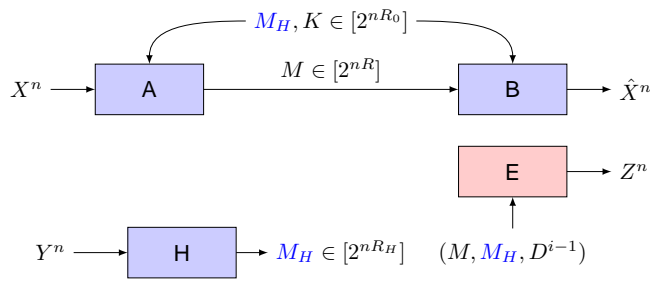
 
 In our max-min formulation, we would like to design encoders \{\textsf{A}, \textsf{H}\} and decoder \Bs to maximize the worst-case payoff with respect to an adversarial eavesdropper  \textsf{E}. This problem subsumes the setting of \cite{rdss}, where there was no helper. However, we provide a full solution only for certain choices of causal disclosure. Traditional approaches to secrecy with a helper can be found in \cite{gund,tand,kc1}. In section VII, we present additional results such as the case when the link from \As to \Bs is private. This recovers the two-sided helper result of \cite{kc1}.
 
Perhaps the most prominent example of communication aided by a public helper appears in the internet, where the helper might be a service provider or a mail client. A more abstract example is provided by team sports, where the helper publicly coaches players \As and \Bs to outperform \textsf{E}. While insight into the structure of optimal strategic communication in the presence of a public helper might be beneficial, we believe that our study has further merits.

Our achievability proof illustrates the versatility of the likelihood encoder \cite{like}. This stochastic approach to encoding seeks to approximate the operational system distribution by an idealized distribution that is extremely simple to analyze. Due to the presence of a helper, we have to use likelihood encoders \As and \Hs that are derived from different idealized distributions. However, we demonstrate that these encoders can mesh together to obtain the desired overall system performance. Also, it is uncertain whether our most general result can be proven using deterministic encoding. This kind of coding is inspired by distributed channel synthesis \cite{DCS}, and can be traced back to Wyner's original ideas \cite{Wyner}.

%By switching the roles of \Hs and \textsf{A}, we extend a triangular source coding result of \cite{kc1}. It is important to note that this reversal of roles is achieved merely by defining appropriate payoff/distortion functions. 
%By setting the distortion function to be independent of \textsf{E}'s actions, we obtain a source coding result that resolves a long-standing issue \cite{kaspi,haim1}.

This approach avoids lengthy entropic manipulations, which usually accompany a purely equivocation-based approach. We leverage the strength of the total variation distance \cite{coord,DCS} to obtain a general result while avoiding consideration of multiple error cases, which are typical of rate-distortion proofs \cite{nit}. The central ingredient of our achievability proof is a generalized soft-covering lemma \cite{DCS,rdss}.

%Also, this result demonstrates the flexibility of the causal-disclosure approach to secrecy.
By choosing the payoff function to be log-loss, we can recover equivocation-based results with respect to the information $X^n$. Unfortunately, we are unable to recover equivocation-based results with respect to \He's information $Y^n$ or $(X^n,Y^n)$ because our converse proof constrains us to exclude $Y^n$ from the payoff function.  However, our achievability proof readily generalizes to these settings.

In this work, we assume that \textsf{A}, \Hs and \Bs have sufficient local randomness. We provide a precise description of the problem in Section II and present a characterization of the optimal rate-payoff region in Section III. Extensions are discussed in section VII.

\section{Preliminaries and Problem Definition}

\subsection{Notation}

We represent both random variables (only finite alphabets) and probability distribution functions with capital letters, but only letters $P$ and $Q$ are used for the latter. We denote the conditional distribution of the random variable $Y$ given the random variable $X$ by $P_{Y|X}(y|x)$, sometimes abbreviated as $P_{Y|X}$. Also, we use the script letter $\Xcal\ni x$ to denote the alphabet of random variable $X$. The set of probabilities (simplex) on $\Xcal$ is denoted by $\D_\Xcal$. Sequences of random variables $X_1,\ldots,X_n$ are denoted by $X^n$. The set $\setb{1,\ldots,m}$ is denoted by $[m]$, while $[m]_+\defeq\max\setb{0,m}$.

Markov chains are denoted by $X-Y-Z$ implying the factorization $P_{XYZ}=P_{XY}P_{Z|Y}$ while $X \perp Y$ indicates that the random variables $X$ and $Y$ are independent. We define the total variation distance as
\begin{equation}
\normtv{P_X-Q_X}\defeq\frac{1}{2}\sum_x\abs{P(x)-Q(x)}.\label{tv}
\end{equation}\vspace{-.4cm}

\subsection{Problem-Specific Definitions}

The communication system model used throughout is shown in Figure \ref{pstsh}. The transmitting node \As observes an iid source sequence $X^n\sim\prod P_X$, while the helper node \Hs observes correlated side information $Y^n\sim\prod P_{Y|X}$. The sequence $D^n\sim\prod P_{D|XY}$ is causally disclosed to node \textsf{E}. Due to a limitation of our converse argument, we only permit $D=(X,D_{x})$ with $P_{D_{x}|X}$ arbitrary. Nodes \As and \Bs share a secret key $K\in[2^{nR_0}]$, which is uniformly distributed and independent of $(X^n,Y^n,D^n)$.

The helper produces a message $M_H\in[2^{nR_H}]$ based on her information $Y^n$, which she broadcasts to both \As and B. Based on the source $X^n$, secret key $K$ and the helper's message $M_H$, \As transmits a message $M\in[2^{nR}]$ that is received by \Bs and \textsf{E}. On receiving $(M,M_H)$, \Bs and \Es make their moves: in the $i$th step, they play $\X_i$ and $Z_i$ respectively. While \Bs produces $\X_i$ based on $(M_H,M,K)$, \Es produces $Z_i$ based on $(M_H,M)$ and the past $D^{i-1}$. Note that the actions of \As are determined by her information $X^n$.

At each step, the joint actions of the players incur a value $\pi(x,\x,z)$, which represents symbol-wise payoff; the block-average payoff is given by 
\begin{equation}
\frac1n \sum_{i=1}^n \pi(X_i,\X_i,Z_i).
\end{equation}
Due to a pruning argument (see Section V.B) in our converse proof, we are constrained to define payoff to be independent of \He's information $Y^n$. Nevertheless, \Hs plays a role in aiding communication. Players \A, \Hs and \Bs want to cooperatively maximize payoff, while \Es tries to minimize payoff through her actions $Z^n$.

\begin{definition}
An $(n,R_H,R,R_0)$ code consists of encoders $f_H:\Ycal^n\rightarrow [2^{nR_H}]$, $f:[2^{nR_H}]\times\Xcal^n\times [2^{nR_0}]\rightarrow [2^{nR}]$ and a decoder $g: [2^{nR_H}]\times[2^{nR}]\times [2^{nR_0}]\rightarrow \Xh^n$. We permit stochastic encoders $P_{M_H|Y^n}$, $P_{M|X^n,M_H,K}$ and a stochastic decoder $P_{\X^n|M_H,M,K}$.
\end{definition}

Nodes \A, \Hs and \Bs use an $(n,R_H,R,R_0)$ code to coordinate against \textsf{E}. We consider payoff against the worst-case adversary. We assume that \Es knows $P_{XYD}$ and the code in use.

\begin{definition}
\label{defnachievability}
 Fix a distribution $P_{XYD}$ and payoff function $\pi:\Xcal\times\Xh\times\Zcal\rightarrow \R$. We say $(R_H,R,R_0,\Pi)$ is achievable if there exists a sequence of $(n,R_H,R,R_0)$ codes such that
%\begin{itemize}
%  \item Under payoff criterion ${\sf P_1}$ (expected payoff):
  \begin{equation}
    \liminf_{n\rightarrow\infty}\min_{\{P_{Z_i|M,D^{i-1}}\}_{i=1}^n}\E\,\frac1n \sum_{i=1}^n \pi(X_i,\X_i,Z_i)\geq\Pi.\label{pay}
  \end{equation}

\end{definition}
With a refined analysis, our main result can be readily extended to more stringent measures such as probability of assured payoff and symbol-wise minimum payoff \cite{rdss}.

Our result allows incorporation of multiple payoff/distortion functions depending on the players' moves to recover results of interest. By convention, payoffs are to be maximized, while distortion is to be minimized (replace $(-\Pi)$ by $\Pi$ in \eqref{pay}).

\begin{definition}
\label{defnregion}
The rate-payoff region $\Rcal$ is the closure of achievable tuples $(R_H,R,R_0,\Pi)$.
\end{definition}

\section{Main Result}
The characterization of the rate-payoff region is given in terms of the following set. Let $\s$ be the set of tuples $(R_H,R,R_0,\Pi)\in\R^4$ such that

\begin{align}
R_H&\ge I(Y;W),\\
R&\ge I(X;UV|W),\\
R_0&\ge I(D;V|U,W),\\
\Pi&\le \min_{z(\cdot,\cdot)}\E\sq{\pi(X,\X,z(U,W))},
\end{align}
evaluated with respect to any $Q_{DXYUVW\X}$ such that
\begin{align}
&(X,Y,D)\sim P_{XYD},\\
&W-Y-XD,\label{mark1}\\
&DY-XW-UVW-\X,\label{mark}
\end{align}
with cardinality bounds $\abs{\mathcal{W}}\le \abs{\mathcal{X}}\abs{\mathcal{Y}}+6,
\abs{\mathcal{U}}\le \abs{\mathcal{X}}\abs{\mathcal{Y}}\abs{\mathcal{W}}+4,
\abs{\mathcal{V}}\le \abs{\mathcal{X}}\abs{\mathcal{Y}}\abs{\mathcal{W}}\abs{\mathcal{U}}|\mathcal{\X}|+2$.
Also, $D=(X,D_{x})$ with $P_{D_{x}|X}$ arbitrary.

\begin{theorem}
\begin{equation}\Rcal=\s.\end{equation}\label{thm}\end{theorem}
\vspace{-0.6cm}

The rate-payoff region is unchanged if the following additional constraints are imposed:
\begin{itemize}
\item $\setbr{V\perp(X,Y,D,W)}$ \textbf{or} $\setbr{H(U|V)=0}$, and
\item \Bs sees past actions $(X^{i-1},Y^{i-1},Z^{i-1})$ at time $i$.
\end{itemize}
Also, the region is achievable for a general disclosure channel $P_{D|XY}$.

\section{Observations}

Our assumption on the disclosure channel $P_{D|XY}$ is made due to a a limitation of our converse argument. This ensures that the desired Markov chains hold in the converse proof. 
Nevertheless, our result addresses the natural choice $D=X$. Also, the important cases of $D=\es$ and when \textsf{B}'s reconstruction is causally disclosed remain unsolved, although they are solved in the absence of a helper \cite[Theorem 1]{rdss}.

Note that setting $W=\es\Rightarrow R_H=0$ recovers \cite[Theorem 1]{rdss}. The Markov chains in $\s$ imply $R+R_H\ge I(XY;UVW)$. This is similar to the communication rate constraint of \cite{rdss}, where the optimal strategy involved giving away part of the communication to \Ee. In our case, the helper merely aids in this aspect.

Since \textsf{H}'s link is public and she does not see the secret key $K$, \Es obtains her codeword $W^n$. Nodes \As and \Hs then perform the scheme of \cite{rdss} conditioned on this side information. That is, \As proceeds to reveal another codeword $U^n$, while using the secret key to keep $V^n$ secret. However, our construction of the distant encoders \As and \textsf{H} needs to address a technical subtlety discussed in section VI.C.

We now present some special cases of our problem, obtained through appropriate choice of payoff/distortion functions and disclosure $D$.

\subsection{Multiterminal Source Coding}

By considering distortion $\pi=-d(x,\x)$ that is independent of \Ee's actions, we obtain a source coding result that recovers \cite[Theorem 2]{haim1}. The projection of $\Rcal$ onto $(R_H,R,\Pi_1)$ is
\begin{align}
R_H&\ge I(Y;W),\\
R&\ge I(X;\X|W),\\
\Pi_1&\ge \E\sq{d_1(X,\X)},
\end{align}
with $W-Y-X$, $Y-XW-\X$ and other constraints fixed.

Incidentally, a solution to the problem of general distortion $d(x,y,\x)$ was claimed by Kaspi-Berger \cite[Theorem 2.1,C]{kaspi}, and refuted by Permuter et al \cite{haim1,haim2} due to an incomplete converse argument. This gap is echoed by our converse proof, which prevents us from considering payoff with respect to $Y^n$.

Whereas our $(R,R_H)$ region is defined by a union of rectangles, \cite[Theorem 2.1,C]{kaspi} proves that a union of larger pentagonal regions is achievable. This is achieved by binning at the helper. In section VII.B, we provide another example where general distortion renders the problem intractable.

\subsection{Equivocation}

By picking $\pi_1=-d_1(x,\x)$ and $\pi_2$ arbitrary, $\Rcal$ transforms to the set
\begin{align}
R_H&\ge I(Y;W),\\
R&\ge I(X;UV|W),\\
R_0&\ge I(D;V|U,W),\\
\Pi_1&\ge \E\sq{d_1(X,\X)},\\
\Pi_2&\le \min_{z(\cdot,\cdot)}\E\sq{\pi_2(X,\X,z(U,W))},
\end{align}
with the same distributional constraints. When we pick  the log-loss $\pi_2=-\log z(x)$ with causal disclosure $D=X$, where $z(x)\in\D_{\mathcal{X}}$, the second payoff reduces to \Ee's normalized equivocation of $X^n$ i.e.\ $n\inv H(X^n|M_H,M)$ \cite{eq} \cite[Lemma 2]{rdss}. The region $\Rcal$ simplifies to

\begin{align}
R_H&\ge I(Y;W),\\
R&\ge I(X;\X|W),\\
\Pi_1&\ge \E\sq{d_1(X,\X)},\\
\Pi_2&\le H(X|W)-[I(X;\X W)-R_0]_+,
\end{align}
with Markov chains $W-Y-X$ and $Y-WX-\X$ and other constraints fixed. The proof is similar to \cite[Corollary 5]{rdss}.

Note that the choice of $\pi_2$ as log-loss effectively makes \Es a passive adversary, in the sense that we know her best strategy \cite[Lemma 2]{rdss}.

%\subsection{Secure Triangular Source Coding}

%Treating $Y^n$ as the information source to be transmitted to \Bs brings up the topic of triangular source coding \cite{triangle}. Set $d_1=d_1(y,\x)$. Additionally, we provide causal disclosure $Y^{i-1}$ to both \As and \Ee, with \As now producing actions $\Xt_i$. We treat \As as an untrustworthy helper, and would like to maximize her equivocation $n\inv H(Y^n|M_H,X^n)$. Defining a worst-case payoff $\pi_3=-\log \xt(y)$, where $\xt\in\D_\Ycal$, with respect to \As as per \eqref{pay}, we obtain the desired equivocation \cite[Lemma 2]{rdss}.

%Equations \eqref{eq1}-\eqref{eq2} transform into the desired rate-distortion-equivocation region as

%\begin{align}
%R_H&\ge I(Y;W),\\
%R&\ge I(X;U|W),\\
%\Pi_1&\ge \E\sq{d_1(Y,\X)},\\
%\Pi_2&\le H(Y|W)-[I(Y;UW)-R_0]_+,\\
%\Pi_3&\le H(Y|XW),
%\end{align}
%with the same distributional constraints. This result extends \cite[Theorem 12]{kc1} when the decoder has no side information. Setting $R_0=0,X=\es$ recovers their result.

%\begin{align}
%R_H&\ge I(Y;W),\\
%R&\ge 0,\\
%\Pi_1&\ge \E\sq{d_1(Y,\X)},\\
%\Pi_2&\le H(Y|W),\\
%\Pi_3&\le H(Y|W),
%\end{align}
%with $W-Y-X$, $Y-WX-W-\X$ and other constraints fixed.

\subsection{Lossless}

With the stronger results mentioned in section II, we can recover results for secure lossless coding by setting $\pi_1(x,\x,z)=\pi(x,z)$ if $\x=x$ and $-\infty$ otherwise \cite[Cor. 1]{rdss}. We omit them here due to a lack of space.

\section{Converse}
\label{sec:converse}

We may assume that Bob can use decoders $\setb{P_{\X_i|M_H,M,K,X^{i-1},Y^{i-1},Z^{i-1}}}_{i=1}^n$. We consider disclosure $D=X$ for simplicity here. %See \cite{web} for the general proof.

\subsection{Bounds}

Let $(R_H,R,R_0,\Pi)$ be achievable. We shall use the random variable $T$ uniformly distributed on $[n]$, as a time index. We use standard information-theoretic inequalities and the fact that $X^n-M_H-K$:

\begin{IEEEeqnarray}{rCl}
nR_H&\ge& H(M_H)\ge I(M_H;X^n,Y^n)\\
&\ge& \sum_{i=1}^nH(X_i,Y_i)-H(X_i,Y_i|M_H,X^{i-1})\\
&\ge& \sum_{i=1}^n I(Y_i;M_H,X^{i-1})\\
&=&n I(Y_T;M_H,X^{T-1},T),%=n I(Y;W),
\end{IEEEeqnarray}

\begin{IEEEeqnarray}{rCl}
nR&\ge& H(M)\ge H(M|K,M_H)\\
&\ge& I(X^n;M|K,M_H)\\
&=&\sum_{i=1}^n I(X_i;M,K|M_H,X^{i-1})\\
&=& n I(X_T;M,K|M_H,X^{T-1},T),
%&= n I(X;UV|W),
\end{IEEEeqnarray}

\begin{IEEEeqnarray}{rCl}
nR_0&\ge& H(K)\ge H(K|M,M_H)\\
&\ge& I(X^n;K|M,M_H)\\
&=& \sum_{i=1}^n I(X_i;K|M,M_H,X^{i-1})\\
&=& n I(X_T;K|M,M_H,X^{T-1},T),%= n I(D;V|U,W).
\end{IEEEeqnarray}

\begin{IEEEeqnarray}{rCl}
\Pi &\leq& \min_{z(\cdot)} \E\frac1n \sum_{i=1}^n \pi(X_i,\X_i,z(M,M_H,X^{i-1},i))\\
 &=& \min_{z(\cdot)} \E\E[\pi(X_T,\X_T,z(M,M_H,X^{T-1},T))|T]\\
 &=& \min_{z(\cdot)} \E\pi(X_T,\X_T,z(M,M_H,X^{T-1},T)),
% &=& \min_{z(\cdot)}\E\pi(X,\X,z(U,W)),
\end{IEEEeqnarray}
where the arguments are $z(m,m_H,x^{i-1},i)$. The desired expressions are obtained by setting $X=X_T$, $Y=Y_T$, $U=M$, $V=K$ and $W=(M_H,X^{T-1},T)$.

\subsection{Pruning}

Note that the above associations inherit Markov chains $W-Y-X$ and $YX-UVW-\X\iff YXW-UVW-\X$. The second Markov chain differs from \eqref{mark}.  Let the induced joint distribution be
\begin{align}
Q&=Q_{\X}Q_{WUV|\X}Q_{YXW|WUV}\\
&=Q_{\X WUV}Q_{XW|WUV}Q_{Y|XWUV}.
\end{align}
Now, let us construct a distribution that satisfies \eqref{mark},
\begin{equation}P=Q_{\X WUV}Q_{XW|WUV} Q_{Y|XW},\end{equation}
where $Q_{Y|XW}$ is induced by $Q$. Firstly, note that
\begin{equation}\sum_y Q=\sum_y P = Q_{\X WUV X},\end{equation}
so the constraints on $(R,R_0,\Pi)$ don't change.

We have $Q_{XW}=P_{XW}$ from above and $P_{Y|XW}=Q_{Y|XW}$ by construction. Also, $P_{YXW}=Q_{YXW}\Rightarrow P_{YW}=Q_{YW}$ so the constraint on $R_H$ does not change. Since $P$ inherits the Markov chain $W-Y-X$ of $Q$ and satisfies \eqref{mark}, we conclude that we can replace $Q$ with $P$, while keeping the rate-payoff region unchanged.

\subsection{Comment on Converse}

The above modification of $Q$ is required in order to recover the desired Markov relation \eqref{mark}. However, note that the trick alters the marginal distribution $Q_{XYWU\X}$ in general. Unfortunately, this prevents us from considering general payoff $d(x,y,\x,z)$. This also explains why the corresponding source coding problem with general distortion remains unsolved \cite{kaspi,haim1,haim2}.

As an aside, the association of $W$ has operational meaning for our problem. Another possibility is to pair $X^{T-1}$ with $M$, which may be fruitful for general disclosure $D$. Also, note that $V\perp (X,Y,W)$.

\section{Sketch of Achievability}

\subsection{Likelihood Encoder}

Optimal play in zero-sum games is often stochastic. As a result, a stochastic decoder is crucial in our work. On the other hand, it is unknown if deterministic encoding suffices. Once we fix our strategy of play, we look for encoders/decoders that recover an iid distribution on all variables. This is the motivation behind likelihood encoding \cite{like}. With the desired average performance guaranteed, we can add any number of payoff functions and the same analysis will guarantee that good encoders/decoders exist.

\subsection{Codebook Construction}

We consider $D=X$ for simplicity. 
%See \cite{web} for general $D$.
Pick a distribution $Q$ of the form that defines $\s$. Generate the helper's codebook: $2^{n I(Y;W)}$ iid $W^n$ codewords indexed by $M_H\in[2^{nR_H}]$. Conditioned on each $W^n$ codeword, generate $2^{n I(X;U|W)}$ iid $U^n$ codewords, indexed by $(M_H,M)\in[2^{nR_H}]\times[2^{nR}]$. For each $(W^n,U^n,K)$ triple, generate $2^{n I(X;V|U,W)}$ iid $V^n$ codewords, indexed by $(M_H,M,K)\in[2^{nR_H}]\times[2^{nR}]\times[2^{nR}]$.

Note that $W-Y-X$ allows the helper to remotely pick a $W^n$ codeword, while $Y-XW-UVW-\X$ reflects the natural flow of information in our scheme: \As sees $(X^n,W^n)$, while \Bs sees $(U^n,V^n,W^n)$.

In keeping with the converse, we may assume that $V\perp(X,Y,D,W)$. This gives secret key $K$ the natural interpretation of facilitating randomized time-sharing between several $V^n$ codebooks.

\subsection{Idealized Distributions}

Consider the distribution $\bp$ obtained by drawing $(M_H,M,K)$ uniformly and passing the resulting $(U^n,W^n,V^n)$ codewords through the memoryless channel $Q_{XY\X|UVW}$. Note that $Y^n-(X^n,W^n)-(U^n,V^n,W^n)-\X^n$. We define \As and \Bs to be $\bp_{U^nV^nW^n|X^nW^n}$ and $\bp_{\X^n|U^nV^nW^n}$ respectively.

Note that defining \Hs with $\bp$ is problematic because she does not see $X^n$. Consider the distribution $\bpa$ obtained by drawing $M_H$ uniformly and passing the resulting $W^n$ codewords through the memoryless channel $Q_{XY|W}$. Note that $W^n-Y^n-X^n$. We set \Hs to $\bpa_{W^n|Y^n}$.

The technical difficulty rests in reconciling \Hs and \{\A,\B\} to obtain the performance under $\bp$. The soft-covering lemma \cite{DCS} ensures that under the $(R_H,R)$ constraints, the joint distribution induced by our choice of \{\A,\B,\He\} approximates $\bp$ in $\normtv{\cdot}$.

\subsection{Attaining Secrecy}

To combat \Ee, we would like enough $K$ to keep $V^n$ secret i.e.\ $V^n\perp(U^n,W^n)$. The soft-covering lemma \cite[Lemma 4]{rdss} ensures this under the $R_0$ constraint. Moreover, a memoryless channel is simulated \cite{DCS} from $(U^n,W^n)$ to $X^n$, so causal disclosure does not help \Ee. %See \cite{web} for details.

\section{Extensions}

\subsection{Private Link from \As to \B}

One might obtain this by defining a new problem where \Es does not see $(M,K)$. Alternatively, note that setting $R_0\ge R$ in $\s$ ensures that $(U^n,V^n)$ are secret in our scheme. The converse arguments are identical. For disclosure $D=X$ and log-loss $\pi_2=-\log z(x)$, where $z(x)\in\D_{\mathcal{X}}$, the second payoff reduces to \Ee's normalized equivocation $n\inv H(X^n|M_H,M)$ \cite{eq} \cite[Lemma 2]{rdss}. The region $\Rcal$ simplifies to

\begin{align}
R_H&\ge I(Y;W),\\
R&\ge I(X;\X|W),\\
\Pi_1&\ge \E\sq{d_1(X,\X)},\\
\Pi_2&\le H(X|W),
\end{align}

with $W-Y-X$, $Y-XW-\X$ and other constraints fixed. This recovers \cite[Theorem 4]{kc1} of Kittichokechai et al.

%\subsection{General Causal Disclosure}

%The important cases of $D=\es$ and when \textsf{B}'s reconstruction is causally disclosed remain unsolved, although they are solved in the absence of a helper \cite[Theorem 1]{rdss}. %The difficulty lies in the converse.

\subsection{Private Side Information}

Consider a problem without \He. When the link from \As to \Bs is public and they share uncoded side information $Y^n$ unseen by \Ee, causal disclosure $D$ leads to a peculiar phenomenon.

% D=X has not been solved for lossless comm. of N; X = N+Y

For concreteness, assume $X=Y\oplus D$ (addition in a finite field), with $Y\perp D$ and $H(Y)\le H(D)$. Let payoff be $\pi=1_\setb{x\ne z}$, the Hamming distance between $X$ and \Ee's reconstruction. Under this model, \As knows $(X^n,Y^n,D^n)$, while \Bs sees $Y^n$. %Note that $H(Y)\le H(X)$.

For lossless communication of $X^n$, the scheme with best-known performance is for \As to send a random enumeration of $D^n$ conditioned on $Y^n$, at rate $H(D)=H(X|Y)$. Given the message $M$, \Es narrows down $D^n$ to a set of size $2^{nH(Y)}$. Since a $2^{-k H(D)}$ fraction of the typical set \cite{Cover} of $D^n$ sequences agrees with causal disclosure $d^k$, \Es learns $D^n$ exactly for times $k>\frac{H(Y)}{H(D)}n$, as $n\to\infty$.

Also, when $k<\frac{H(Y)}{H(D)}n\iff nH(Y)>kH(D)$, the block $D^k$ is concealed from \Es because the random enumeration acts as an unstructured one-time pad \cite{Cuff3}, as $k\to\infty$.
%the soft-covering lemma \cite{DCS} implies that the block $D^k$ appears iid to \Es as $n\to\infty$, 
Hence, causal disclosure does not help. When $Y\sim \mbox{Bern}(p)$ ($0\le p\le1/2$) and $D\sim \mbox{Bern}(1/2)$, \Es incurs an average payoff of $\frac{H(Y)}{H(D)}(1/2)+(1-\frac{H(Y)}{H(D)})(p)=p+h(p)(1/2-p)$, where $h(\cdot)$ is the binary entropy function. It is unknown whether this scheme is optimal.

% note that the rate requirement in soft-covering is $I(D;X|Y)=H(D|Y)=H(D)$
% we are given iid Y^n, and the channel is from (X^n,Y^n) to D^n
% There are at least 2^nH(Y) candidate X^n sequences, one for each Y^n
% addition is a memoryless channel

% A covers (X^n,Y^n) with D^n and sends the bin index (binned wrt Y^n at rate H(D|Y)=H(D))
% Given the bin index, we want the size of the Y^n "codebook" to be larger than H(D|Y)=H(D)
% The soft-covering lemma will then yield perfect secrecy for k < H(Y)/H(D) n
% The random enumeration (binning at rate 0) acts as an unstructured one-time pad

Remarkably, the same problem for payoff $\pi=1_\setb{d\ne z}$ is solved by our result and \cite[Theorem 1]{rdss}. However, the problem changes dramatically when the side information $Y$ is introduced into the payoff function. This example also illustrates that an equivocation-based approach is indifferent to securing just a fraction $X^{\br{\frac{H(Y)}{H(D)}}n}$ of the source sequence versus partially securing the whole sequence.

%% Appendix:
%% If needed a single appendix is created by
%\appendix
%% If several appendices are needed, then the command
%\appendices
%% in combination with further \section-commands can be used.

%% Use \section* for acknowledgement
\section*{Acknowledgment}

The authors would like to thank Curt Schieler for insightful discussions. This work is supported by the National Science Foundation (grant CCF-1116013) and the Air Force Office of Scientific Research (grant FA9550-12-1-0196).\vspace{-.1cm}

%% References:
%% We recommend the usage of BibTeX:
%%
%\bibliographystyle{IEEEtran}
%\bibliography{definitions,bibliofile}
%%
%% where we here have assume the existence of the files
%% definitions.bib and bibliofile.bib.
%% BibTeX documentation can be obtained at:
%% http://www.ctan.org/tex-archive/biblio/bibtex/contrib/doc/
%%
%%
%%
%% Or manual references (pay attention to consistency!):
%\begin{thebibliography}{1}
%\bibitem{shannon1948}
%  C.~E. Shannon, ``A mathematical theory of communication,''
%  \emph{Bell System Techn. J.}, vol.~27, pp. 379--423 and 623--656,
%  Jul. and Oct. 1948. 
%\end{thebibliography}

\bibliographystyle{ieeetr}

\bibliography{helper_isit}

\end{document}